\documentstyle[12pt]{article}
\newcommand{\text}{\rm}
\begin{document}

\title{{\bf Bose-Einstein condensation and superfluidity of a weakly-interacting
photon gas in a nonlinear Fabry-Perot cavity}}
\author{{\bf A. Tanzini and S.P. Sorella } \and \vspace{2mm} \\
UERJ, Universidade do Estado do Rio de Janeiro\\
Departamento de F\'\i sica Te\'orica\\
Instituto de F\'{\i}sica\\
Rua S\~ao Francisco Xavier, 524\\
20550-013, Maracan\~{a}, Rio de Janeiro, Brazil \\
\vspace{2mm} \and {\bf UERJ/DFT-02/99}\vspace{2mm}\newline}
\maketitle

\begin{abstract}
A field theoretical framework for the recently proposed photon condensation
effect in a nonlinear Fabry-Perot cavity is discussed. The dynamics of the
photon gas turns out to be described by an effective 2D Hamiltonian of a
complex massive scalar field. Finite size effects turn out to be relevant
for the existence of the photon condensate.

\setcounter{page}{0}\thispagestyle{empty}
\end{abstract}




\vfill\newpage\ \makeatother
\renewcommand{\theequation}{\thesection.\arabic{equation}}

\section{Introduction}

This work originates from a very recent experimental proposal \cite{exp} in
order to detect a Bose-Einstein condensation and a new superfluid state of
light. The apparatus \cite{exp} consists basically of a planar Fabry-Perot
cavity filled with a nonlinear polarizable medium responsible for an
effective repulsive short-range four-photon interaction. The mirrors of the
cavity have a low but finite transmittivity, allowing photons of an incident
laser beam to enter and leave the cavity, so that a steady-state condition
is achieved after many photon-photon collisions. Following \cite{exp}, the
weak interaction between the photons can be viewed as a four-photon term
arising from a repulsive pairwise interaction provided by a self-defocusing
Kerr nonlinear medium inside the Fabry-Perot cavity \cite{int}. Moreover,
due to the boundary conditions required by the Fabry-Perot, one is led with
an effective two-dimensional nonrelativistic massive weakly-interacting
photon gas confined in the cavity. The nonrelativistic regime is due to the
paraxial propagation of the light inside the resonator \cite{exp,int}. This
effective 2D weakly-interacting photon gas displays a Bogoliubov type
dispersion relation, suggesting the existence of a Bose-Einstein photon
condensate and of a possible superfluid state of light. In particular, one
of the tasks of the experiment is to investigate the existence of the
sound-like waves corresponding to the collective phonon excitations in the
photon superfluid, which, according to \cite{exp}, should propagate with a
velocity whose value $v_c$ is a few thousandths of the vacuum speed of
light, $v_c=4.2\times 10^7cm/s$. A follow-up experiment is being planned 
\cite{exp} in order to demonstrate that the sound wave velocity $v_c$ is a
critical velocity for the photon (super)fluid, corresponding to the
existence of persistent currents.

It is useful now to spend a few words on the meaning of the Bose-Einstein
condensation for this 2D photon gas, as is well known that there are quite
severe restrictions on the existence of a Bose-Einstein condensation in
momentum space for a weakly-interacting 2D gas in the thermodynamic limit
for any nonvanishing temperature \cite{nbec, mul,mul1}.

\noindent
We underline first of all that the system realized by the experimental
set-up of \cite{exp} is a zero-temperature Bose gas. We know that for an
ideal Bose gas a macroscopic number $N_0$ of particles will condense
occupying the zero-momentum state; however the presence of interactions
could strongly modify this picture, in such a way that the presence of the
condensate is no longer so obvious \cite{ab,be}. Nevertheless, in our case
the interaction is very weak, and we may think that $N_0$ remains a
macroscopic number. Moreover, the presence of a nonvanishing interaction is
crucial in order to imply a redefinition of the spectrum of the excitations
which gives rise to the Bogoliubov dispersion relation.

\noindent
Another remarkable issue is the fact that the photon gas ought to be
considered here as a genuine finite-sized system, as the whole apparatus
possesses a finite small volume ({\it i.e. }the cavity volume) and the
average number $N$ of photons inside the cavity is kept finite as well.
Also, the Fabry-Perot boundary conditions play a crucial role in order to
provide an effective mass for the photons \cite{exp}, which turns out to be
proportional to the inverse of the separation length $L$ between the mirrors
of the cavity. The situation looks very close to that of trapped Bose gases 
\cite{trap1,trap2,trap3}, for which the Bose-Einstein condensation has been
observed even in the case of 1D dimensionally reduced systems \cite{vdk,str}%
. For these inhomogeneous finite-sized systems the existence of a
Bose-Einstein condensation is not, strictly speaking, a phase transition 
\cite{str,wb,mul,mul1}. Rather, it is a direct consequence of the
experimental evidence \cite{trap1,trap2,trap3,str} of a macroscopic
occupation of the lowest state. It is worth remarking that the number $N_0$
of photons occupying the lowest state has been estimated \cite{exp} to be of
the order of $N_0=8\times 10^{11}$.

The aim of the present work is to propose and analyse a possible theoretical
set up. We shall be able to show that the effective 2D weakly-interacting
massive photon gas can be actually obtained by a four-photon $QED$-inspired
Hamiltonian, once the gauge freedom and the Fabry-Perot boundary conditions
have been properly taken into account.

\section{The Effective Hamiltonian}

As the starting $QED$-inspired effective Hamiltonian describing a weak
repulsive four-photon interaction we take the following gauge invariant
expression

\begin{equation}
H_{{\rm eff}}=\int_Vd^3x\left( \frac 12\left( \overrightarrow{E^2}+%
\overrightarrow{B^2}\right) +\frac \lambda 4\left( A_\mu ^TA^{T\mu }\right)
^2\right) \;,  \label{heff}
\end{equation}
$V$ being the volume\footnote{%
As we shall see in the next section, the boundary conditions required by the
Fabry-Perot cavity do allow for usual integration by parts.} of the
Fabry-Perot cavity. The coupling constant $\lambda $ is positive and $A_\mu
^T$ stand for the transverse gauge invariant components of the gauge field, 
{\it i.e.}

\begin{equation}
A_\mu ^T=(g_{\mu \nu }-\frac{\partial _\mu \partial _\nu }{\partial ^2}%
)A^\nu =\frac 1{\partial ^2}\partial ^\nu F_{\nu \mu }\;,\;  \label{tr}
\end{equation}
where $g_{\mu \nu }={\rm diag}(+,-,-,-)$ is the flat Minkowski metric and

\begin{equation}
F_{\mu \nu }=\partial _\mu A_\nu -\partial _\nu A_\mu \;.  \label{fmn}
\end{equation}
In order to motivate the choice of the Hamiltonian $\left( {\rm {\ref{heff}}}%
\right) $ we underline that, according to \cite{int}, the field propagation
inside the cavity in the paraxial approximation is described by a nonlinear
Schr\"{o}dinger equation. The effective Hamiltonian $\left( {\rm {\ref{heff}}%
}\right) $ can then be obtained by requiring that the Heisenberg equations
of motion for the field reproduce the nonlinear Schr\"{o}dinger equation in
the semiclassical limit \cite{int}.

Although not needed, it is worth remarking that expression $\left( {\rm {\ref
{heff}}}\right) $ can be immediately generalized in a gauge invariant way to
a typical two-body interaction, namely

\begin{equation}
H_{{\rm eff}}=\int_Vd^3x\frac 12\left( \overrightarrow{E^2}+\overrightarrow{%
B^2}\right) +\frac 14\int_Vd^3xd^3y\;\left( A^TA^T\right) (x)U(x-y)\left(
A^TA^T\right) (y)\;,  \label{t-b}
\end{equation}
for some short-range repulsive potential $U(x-y).$

Being interested in the analysis of the ground state of the Hamiltonian $%
\left( {\rm {\ref{heff}}}\right) $, we shall work in the static situation in
which all fields are assumed to be time-independent. Accordingly, we shall
make use of the so called temporal gauge

\begin{eqnarray}
A_0 &=&0\;,  \label{t-g} \\
\overrightarrow{A} &=&\overrightarrow{A}(\overrightarrow{x})\;,\;  \nonumber
\end{eqnarray}
which implies

\begin{eqnarray}
A_0^T &=&0\;,  \label{sptr} \\
A_i^T &=&\left( \delta _{ij}-\frac{\partial _i\partial _j}{\nabla ^2}\right)
A_j=\frac 1{\nabla ^2}\partial _jF_{ji}\;,\;\;\;\;\;i,j=1,2,3\;,  \nonumber
\end{eqnarray}
where $1/\nabla ^2$ is the Green's function of the three-dimensional
laplacian,

\begin{equation}
\frac 1{\nabla ^2}=-\frac 1{4\pi \left| \overrightarrow{x}-\overrightarrow{y}%
\right| }\;.  \label{lap}
\end{equation}
For the Hamiltonian $\left( {\rm {\ref{heff}}}\right) $ we get

\begin{equation}
H_{{\rm eff}}=\int d^3x\left( \frac 14F_{ij}F^{ij}+\frac \lambda 4\left(
A_i^TA_i^T\right) ^2\right) \;.  \label{h1}
\end{equation}
Obviously, expression $\left( {\rm {\ref{h1}}}\right) $ is left invariant by
the time-independent gauge transformations, {\it i.e.}

\begin{equation}
\delta A_i=\partial _i\eta (\overrightarrow{x})\;.  \label{ti-gt}
\end{equation}
This spatial type of gauge invariance can be fixed by imposing the axial
gauge condition

\begin{equation}
A_3(\overrightarrow{x})=0\;,  \label{ax-cond}
\end{equation}
naturally suggested by the geometry of the Fabry-Perot cavity. Here the $z$%
-axis is chosen to be coincident with the direction of propagation of the
laser beam incident on the Fabry-Perot. The mirrors of the cavity lie in the
transverse $xy$ plane.

\noindent
However, as is well known, condition $\left( {\rm {\ref{ax-cond}}}\right) $
does not fix completely the gauge freedom and allows for a further residual
local invariance, corresponding to the gauge transformations on the plane $%
xy $ ortogonal to the $z$-axis$.$ In fact, owing to the equation $\left( 
{\rm {\ref{ax-cond}}}\right) ,$ for the field strength $F_{ij}$ we get

\begin{eqnarray}
F_{a3} &=&-\partial _3A_a\;,  \label{res-g} \\
F_{ab} &=&\partial _aA_b-\partial _bA_a\;,\;\;\;\;\;\;a,b=1,2\;.  \nonumber
\end{eqnarray}
It is apparent then that the components of $F_{ij}$ are left invariant by
the following $z$-independent transformations

\begin{equation}
\delta A_a=\partial _a\eta (x_{\bot })\;,  \label{tr-g}
\end{equation}
where $\overrightarrow{x}_{\bot }=(x,y)$. This further residual gauge
invariance $\left( {\rm {\ref{tr-g}}}\right) $ can be fixed by requiring the
additional condition

\begin{equation}
\partial _aA_a=0\;,  \label{res-cond}
\end{equation}
from which it follows that the two gauge fields $A_a$ can be identified with
their transverse components. Finally, for the fully gauge fixed effective
Hamiltonian we obtain

\begin{eqnarray}
H_{{\rm eff}} &=&\int d^2x_{\bot }dz\left( \frac 12F_{3a}F^{3a}+\frac
14F_{ab}F^{ab}+\frac \lambda 4\left( A_aA_a\right) ^2\right) \;  \label{h-2}
\\
&=&\int d^2x_{\bot }dz\left( \frac 12(\partial _3A_a)(\partial _3A_a)-\frac
12A_a\nabla _{\bot }^2A_a+\frac \lambda 4\left( A_aA_a\right) ^2\right) \;, 
\nonumber
\end{eqnarray}
where $\nabla _{\bot }^2=\partial _a\partial _a$ is the two-dimensional
laplacian. This Hamiltonian will be the starting point for the analysis of
the spectrum of the excitations of the weakly coupled photon gas.

\section{The spectrum of the excitations}

In order to analyse the spectrum of the Hamiltonian $\left( {\rm {\ref{h-2}}}%
\right) $, we have first to properly take into account the boundary
conditions of the problem. These require the {\it vanishing} of the fields
at the reflecting surfaces of the mirrors of the Fabry-Perot cavity, {\it %
i.e.}

\begin{equation}
A_a(x_{\bot },z)=\frac 1{\sqrt{L}}\widetilde{A}_a(x_{\bot })\sin (\frac \pi
Ln_0z)\;,  \label{bc}
\end{equation}
where $L$ is the distance between the mirrors and where the fixed integer $%
n_0$ is related to the frequency $\omega $ of the laser beam incident on the
cavity through $n_0\pi /L=\omega $. It should be pointed out that the form
of the field $\left( {\rm {\ref{bc}}}\right) $ requires the assumption that
the spacing between the modes of the cavity is so large that only one
longitudinal mode is excited by the laser beam \cite{int}. Concerning now
the $xy$ plane, periodic boundary conditions will be assumed.

Inserting equation $\left( {\rm {\ref{bc}}}\right) $ in the expression $%
\left( {\rm {\ref{h-2}}}\right) $ and performing the integration over the $z$%
-axis, we easily get the following $2D$ dimensionally reduced effective
Hamiltonian

\begin{equation}
H_{{\rm eff}}=\int d^2x_{\bot }\left( -\frac 14\widetilde{A}_a\nabla _{\bot
}^2\widetilde{A}_a+\frac{m^2}4\widetilde{A}_a\widetilde{A}_a+\frac
3{32L}\lambda \left( \widetilde{A}_a\widetilde{A}_a\right) ^2\right) \;,
\label{h-3}
\end{equation}
where $m=n_0\pi /L=\omega $ is the effective mass of the photon gas confined
in the Fabry-Perot cavity. It is worth mentioning that the paraxial
approximation guarantees that the photons have a finite effective mass also
when tunneling effects due to the low but finite transmittivity of the
mirrors are taken into account \cite{exp}. Setting

\begin{eqnarray}
\widetilde{A}_1 &=&(\varphi +\varphi ^{\dagger })\;,  \label{comp} \\
\widetilde{A}_2 &=&-i(\varphi -\varphi ^{\dagger })\;,  \nonumber
\end{eqnarray}
we obtain the final form of the four-photon Hamiltonian

\begin{equation}
H_{{\rm eff}}=\int d^2x_{\bot }\left( -\varphi ^{\dagger }\nabla _{\bot
}^2\varphi +m^2\varphi ^{\dagger }\varphi +\frac 3{2L}\lambda \left( \varphi
^{\dagger }\varphi \right) ^2\right) \;,  \label{h-f}
\end{equation}
describing an effective weakly-interacting massive 2D photon gas. This
Hamiltonian displays a $U(1)$ global phase symmetry, which follows from the $%
O(2)$ rotational invariance in the $xy$ plane of the dimensionally reduced $%
2D$ effective Hamiltonian $\left( {\rm {\ref{h-3}}}\right) $. Moreover, for
a paraxial propagation of the light inside the cavity \cite{exp}, we have $%
p_{\bot }=\sqrt{p_1^2+p_2^2}\ll m$, so that the photon gas is in fact
nonrelativistic. In order to obtain the spectrum of the Hamiltonian we
expand the fields $\varphi ,$ $\varphi ^{\dagger }$ in Fourier modes

\begin{eqnarray}
\varphi &=&\frac 1{\sqrt{A}}\sum_pa_pe^{i\overrightarrow{p}_{\bot }\cdot 
\overrightarrow{x}_{\bot }}\;,  \label{fm} \\
\varphi ^{\dagger } &=&\frac 1{\sqrt{A}}\sum_pa_p^{\dagger }e^{-i%
\overrightarrow{p}_{\bot }\cdot \overrightarrow{x}_{\bot }}\;,  \nonumber
\end{eqnarray}
where $A$ is the available area of the cavity in the transverse plane $xy.$
Thus

\begin{equation}
H_{{\rm eff}}=\sum_p\epsilon _pa_p^{\dagger }a_p+\frac{3\lambda }2\frac
1V\sum_{p_1+p_2=p_3+p_4}a_{p_1}^{\dagger }a_{p_2}^{\dagger }a_{p_3}a_{p_4}\;,
\label{h-ff}
\end{equation}
where $V=AL$ is the volume of the cavity and

\begin{equation}
\epsilon _p=\sqrt{\overrightarrow{p}_{\bot }^2+m^2}\;.  \label{exc}
\end{equation}
Of course, expression $\left( {\rm {\ref{h-ff}}}\right) $ is nothing but the
starting Hamiltonian of \cite{exp}. Now we can proceed with the standard
analysis of the weakly-interacting Bose gas within the Bogoliubov
approximation \cite{exp,wb}. Assuming that the number $N_0$ of excitations
occupying the zero-momentum state is macroscopic\footnote{%
We recall here that $N_0$ is of the order of $8\times 10^{11}$, so that the
condition $N_0\gg 1$ is in fact verified.}, {\it i.e. }$N_0\gg 1$, and
neglecting higher order interaction terms above the condensate, the
Hamiltonian $\left( {\rm {\ref{h-ff}}}\right) $ is diagonalized by means of
a Bogoliubov transformation

\begin{eqnarray}
a_p &=&u_p\alpha _p+v_p\alpha _{-p}^{\dagger }\;,  \label{b-t} \\
a_p^{\dagger } &=&u_p\alpha _p^{\dagger }+v_p\alpha _{-p}\;,  \nonumber
\end{eqnarray}
with

\[
u_p^2-v_p^2=1\;. 
\]
The resulting spectrum is easily worked out and turns out to be given by

\begin{equation}
H_{{\rm eff}}=\sum_{p\neq 0}\widetilde{\epsilon }_p\alpha _p^{\dagger
}\alpha _p+\;E_0\;,  \label{h-d}
\end{equation}
with

\begin{equation}
\widetilde{\epsilon _p}=\left[ \;\left( \epsilon _p-m+3\lambda \frac
NV\right) ^2-\left( 3\lambda \frac NV\right) ^2\right] ^{1/2}\;,
\label{bexc}
\end{equation}
$N$ being the total average number of photons in the cavity. As is well
known, expression $\left( {\rm {\ref{bexc}}}\right) $ provides a
nonvanishing critical velocity for the phonon sound-like waves

\begin{equation}
v_c=\lim_{p\rightarrow 0}\frac{\widetilde{\epsilon _p}}p=\sqrt{\frac{%
3\lambda }m\frac NV}\;.  \label{vc}
\end{equation}
Let us finally display the time evolution of the fields in the Bogoliubov
approximation $\left( {\rm {\ref{h-d}}}\right) $. Making use of the relations

\begin{eqnarray}
u_p^2+v_p^2 &=&\frac{\left( \epsilon _p-m+3\lambda \frac NV\right) }{%
\widetilde{\epsilon _p}}\;,  \label{uv-r} \\
2u_pv_p &=&-3\frac NV\frac \lambda {\widetilde{\epsilon _p}}\;,  \nonumber
\end{eqnarray}
for the time evolution of the mode $a_p$ we obtain

\begin{eqnarray}
i\frac{\partial a_p(t)}{\partial t} &=&\left( \epsilon _p-m\right)
a_p(t)+3\lambda \frac NV\left( a_p(t)+a_{-p}^{\dagger }(t)\right) \;,
\label{t-e} \\
a_p(t) &=&e^{iH_{{\rm eff}}t}a_pe^{-iH_{{\rm eff}}t}\;,  \nonumber
\end{eqnarray}
in agreement with the classical nonlinear argument of \cite{exp}.

\section{Conclusion}

In this work we have derived a possible theoretical set up for the photon
condensation effect proposed in \cite{exp}, starting from a nonlinear $QED-$%
inspired Hamiltonian. In particular, we have shown that the dynamics of the
photon gas inside the Fabry-Perot cavity filled with a nonlinear polarizable
medium is described in terms of a $2D$ dimensionally reduced effective
Hamiltonian of a massive complex scalar field. This implies that the
transverse dimensions of the cavity should play a crucial role in the photon
condensation. In fact, according to our field theory description, if one
takes as possible thermodynamic limit\footnote{%
We remind that $A$ is the available area of the cavity in the transverse
plane $xy$.} $N,$ $A\rightarrow \infty ,$ $N/A=$const., the lowest lying
phonon of the Bogoliubov spectrum would play the role of the Goldstone boson
corresponding to the spontaneous breaking of the $U(1)$ symmetry of the
Hamiltonian (\ref{h-f}). However, it is well known that spontaneous symmetry
breaking {\it cannot} take place in two-dimensions, due to the infrared
divergencies associated to massless scalar fields \cite{col}.

Therefore, as in the case of trapped Bose gases, the photon gas has to be
considered as a genuine finite-sized system, the transverse dimensions of
the cavity providing a natural infrared cut-off\footnote{%
Notice in fact that finite size effects are rather relevant in the
condensation of 2D dimensionally reduced atomic gases, for very similar
reasons: long wavelenght phonons destabilize the long range order of the
condensate \cite{mul1}.}.

Thus, we expect that these finite size conditions should be actually
realized in the experimental framework in order to observe the photon
condensate. The role played by the finiteness of the system is under
investigation. In this context, it would be also very interesting to study
the possibility of the existence of a Kosterlitz-Thouless phase transition
for the $2D$ photon (super)fluid.

\section{Acknowledgements}

We are very grateful to R.Y. Chiao for valuable suggestions and comments. We
are indebted to P. Di Porto, D. Barci and E. Fraga for fruitful discussions.
The Conselho Nacional de Desenvolvimento Cient\'{\i}fico e Tecnol\'{o}gico
CNPq-Brazil, the Funda{\c {c}}{\~{a}}o de Amparo {\`{a}} Pesquisa do Estado
do Rio de Janeiro (Faperj) and the SR2-UERJ are acknowledged for the
financial support.

\end{document}